\documentclass[12pt]{article}
\usepackage[T1]{fontenc}
\usepackage[utf8]{inputenc}

\usepackage{xcolor}
\usepackage{amsmath}
\usepackage{bm}
\usepackage{relsize}
\usepackage{comment}
\usepackage{tgtermes}
\usepackage{graphicx}
\usepackage[UKenglish]{babel}
\usepackage{microtype}                      
\usepackage{xcolor} 
\usepackage{hyperref}                   	
\usepackage{lipsum}
\usepackage{fancyhdr}
\usepackage{picture} 
\hypersetup{								
	colorlinks=true,      		            
	linkcolor=blue,                        	
	filecolor=blue,                     	
	citecolor=blue,    	        			
	urlcolor=blue,                 	   		
	pdffitwindow=false,               	 	
	pdfstartview={FitH},               		
	pdfauthor={Huw Price},           		
}

\usepackage{braket}
\usepackage{physics}
\usepackage[title]{appendix}

\newcommand{\W}{$\lor\hspace{-2pt}\lor$}
\newcommand{\V}{$\lor$}

\begin{document}               
\title{Entanglement Swapping and Action at a Distance}%
\author{Huw Price\thanks{Trinity College, Cambridge CB2 1TQ, UK; email \href{mailto:hp331@cam.ac.uk}{hp331@cam.ac.uk}.} {\ and} Ken Wharton\thanks{Department of Physics and Astronomy, San Jos\'{e} State University, San Jos\'{e}, CA 95192-0106, USA; email \href{mailto:kenneth.wharton@sjsu.edu}{kenneth.wharton@sjsu.edu}.}}
\date{September 2021}
\maketitle\thispagestyle{empty}

\begin{abstract}
\noindent A 2015 experiment by Hanson and Delft colleagues provided further confirmation that the quantum world violates the Bell inequalities, being the first Bell test to close two known experimental loopholes simultaneously. The experiment was also taken to provide new evidence of `spooky action at a distance'. Here we argue for caution about the latter claim. The  Delft experiment relies on entanglement swapping, and our main claim is that this geometry introduces an additional loophole in the argument from violation of the Bell inequalities to action at a distance: the apparent action at a distance may be an artifact of `collider bias'. In the absence of retrocausality, the sensitivity of such experiments to this   `Collider Loophole'  (CL) depends on the temporal relation between the entanglement swapping measurement C and the two measurements A and B between which we seek to infer a  causal connection. CL looms large if the C is in the future of A and B, but not if C is in the past. The Delft experiment itself is the intermediate case, in which the separation is spacelike. We argue that this leaves it vulnerable to CL, unable to establish conclusively that it avoids it. 
An Appendix discusses the implications of permitting retrocausality for the issue of causal influence across entanglement swapping measurements.\\

\end{abstract}

\newpage\setcounter{page}{1}
\section{Introduction}

\noindent In 2015, Ronald Hanson and colleagues at Delft reported  the first of an important new class of `Bell tests' -- i.e., experimental confirmations that the quantum world violates the Bell inequalities \cite{Hensen15}. The Delft experiment was the first Bell test to close {both} of two well-known experimental loopholes, the so-called \textit{detection} and \textit{locality} loopholes. Previous Bell tests had closed one or the other, but not both simultaneously. As \cite[3]{Hensen15}  noted, their experiment exploited `[a]n elegant approach for realizing a loophole-free setup \ldots\ proposed by Bell himself'. The same approach has now been taken by several other experiments \cite{Giustina15,Shalm15,Rosenfeld17}. 

The Delft experiment was widely hailed as a further important confirmation of quantum `action at a distance' (AAD). Media coverage  presented the experiment as good news for AAD and hence as bad news for Einstein: ``The most rigorous test of quantum theory ever carried out has confirmed that the `spooky action-at-a-distance' that [Einstein] famously hated \ldots\ is an inherent part of the quantum world,'' as a report in \emph{Nature} put it \cite{Merali15}.

Here we argue that this conclusion needs a large caveat. The Delft experiment does, as claimed, do a convincing job of closing experimental loopholes in the project of showing that nature violates the Bell inequalities. But the step from violation of the Bell inequalities to AAD involves a sensitivity to experimental geometry that has not previously been recognised, so far as we are aware. Our goal in this paper is to call attention to this issue, using the Delft experiment and some variants of it as a framework for our discussion.

The issue arises because, unlike previous Bell tests, such experiments make use of  \textit{entanglement swapping.} Their spacetime geometry thus has a \W\ shape, rather than the \V\ shape of previous experiments. The central vertex of the \W\ is a measurement whose result (if suitable) confirms an entanglement between the particles measured at the outer vertices. We will show that this geometry may permit an alternative explanation of the observed Bell correlations across the \W, not requiring AAD. The apparent  causal influence may be a selection artifact, of a kind familiar in the causal modelling literature, rendered possible in this case by the role of the central measurement. 
 
 Specifically, the central measurement is a `collider' in causal modelling terms, with the result that apparent causal influence across the \W\ may be an artifact of so-called `collider bias'. (We explain these terms below.) 
 This means that such an experiment may  provide no evidence of AAD in its own case -- i.e., across the \W\ as a whole\footnote{We emphasise that the argument is not a challenge to the claim that there is AAD \textit{within} the two \V-shaped wings of the experiment. As we will argue, however, this need not add up to AAD across the \W\ as a whole.} -- despite confirming predictions that seem to mandate AAD in other geometries. We will call this the \textit{Collider Loophole} (CL).
 
 We will show that under standard assumptions, vulnerability of such experiments to CL depends on the spacetime location of the central vertex of the \W\ with respect to the outer vertices. Cross-\W\ AAD is highly questionable if the central vertex lies in the absolute future of the outer vertices, but not if it lies in their absolute past. Interestingly, the Delft experiment itself is the intermediate case, in which this separation is spacelike. We will argue that in the light of this, the case for AAD across the \W\ in the Delft experiment is weaker than in the subsequent similar experiments, in which the central vertex lies in the overlap of the past light cones of the outer vertices.

 The sensitivity of CL to the spacetime location of the central vertex depends on the assumption that there is no retrocausality in the systems in question. This may seem uncontroversial, but it is has been challenged in this context, an option that has been held to provide a different reason for questioning the inference from violation of the Bell inequalities to AAD. Retrocausal models allow causality `across the \V' in conventional \V-shaped Bell experiments, but take it to be indirect: the causal influence is said to take a zig zag path, along the two arms of the \V. The result is spacelike causality, without direct AAD. The claimed advantage is that by keeping direct causal influences within the light cones, such models may be easier than conventional models to reconcile with special relativity; see \cite{
FriedrichEvans19, WhartonArgaman20, NorsenPrice21} for recent discussions. 

In the body of the paper we ignore this option. Indeed, a \textit{No Retrocausality Assumption} (NoRA) will play a crucial role in our argument, supporting the case for the existence of the Collider Loophole in some \W\ geometries but not others. In the Appendix we ask what difference it makes to our conclusions if we abandon NoRA (i.e., permit retrocausality). We argue that it would eliminate this difference between \W\ geometries -- that  being an advantage -- but in such a way as to make all of them vulnerable to CL. We note that this suggests new responses to some objections to retrocausal approaches, including an objection based on the possibility of iterated zig zag causality. These issues provide a further motivation for the main project of the paper, that of exploring the question of causal influence across \W\ geometries.

The paper goes like this. \S2 deals with preliminaries. We introduce entanglement swapping, and a variant of it known as \textit{delayed choice} entanglement swapping (DCES). We summarise recent discussion in the literature about the ontological status of the entanglement produced by DCES. In particular, we explain the common view that DCES does not create `genuine' entanglement, and that the appearance that it does so is an artifact of post-selection. We also introduce the notions of colliders and  collider bias from the causal modelling literature.

In \S3 we describe the Delft experiment, and then apply the lessons of \S2 to a hypothetical variant of it, conducted with DCES. We show that in this DCES variant, it would be highly questionable whether violation of the Bell inequalities would reflect any real AAD across the \W\ geometry of the experiment as a whole. The alternative is that the Bell correlations are an artifact of collider bias. We give two reasons for thinking that this is the true explanation of the cross-\W\ correlations in the DCES case. First, the central measurement is a collider, which immediately puts the possibility of collider bias on the table. Second, the experimental correlations fail two intuitive tests for the existence of a genuine (cross-\W) causal connection. We argue that taken together, these factors provide strong reasons to doubt cross-\W\ AAD in the DCES case.\footnote{Readers may be familiar with a different concern about post-selection in Bell experiments, a concern applicable  in conventional \V-shaped experimental geometries. We emphasise that the issue we raise here is different, and arises specifically from the use of entanglement swapping. See footnote \ref{footnote_1} 
below for a further comment on this distinction.\label{footnote_2} (See also \cite{Guido21} for recent relevant work.)}

Up to this point, we will have been speaking about AAD, not `nonlocality' or `nonlocal causation'. This choice is deliberate, and in \S4 we explain why we make it. Bell's own formal notion of \textit{Local Causality} (LC) \cite{Bell90} turns out to require particular care, in these contexts. When selection artifacts are in the offing, it is important to distinguish a version of LC expressed in terms of frequencies in post-selected ensembles from a version (Bell's own) expressed in term of underlying fundamental probabilities. In \S4  we explain the need for this distinction, and then discuss the relation between the failure of cross-\W\ AAD and LC (in both senses).

In \S5 we turn to a second variant of the Delft experiment, in which the entanglement swapping occurs in the past of the measurements at the extremities of the \W\  (i.e., in the overlap of their past light cones). As we will explain, this variant appears to avoid the Collider Loophole.  

In \S6, with these variants as comparisons, we return to the actual Delft experiment.  In this case, as  we noted, the entanglement swapping  measurement occurs at spacelike separation from the two measurements between which the experiment claims to reveal AAD. This makes a difference, but we conclude that although the case for thinking that the experiment fails to close the Collider Loophole is not as straightforward as in the delayed choice version,  the  loophole remains a threat. 

\S7 is a brief conclusion to the main argument. In Appendix A, finally, we return to  the option of admitting retrocausality, and discuss its implications for the question of AAD across the \W-geometry. 

\section{Preliminaries}
\subsection{Entanglement swapping}

We introduce entanglement swapping by following a helpful recent presentation in \cite{Glick19}. 
As Glick puts it:
\begin{quote}
\noindent Entanglement swapping is a procedure in which entanglement
may be ``swapped'' from a pair of jointly measured particles to a pair
of particles lacking common preparation. The technique has
become quite commonplace in experiments involving entanglement and has numerous applications in quantum information
theory. 
 A simple experimental
arrangement is depicted  [in Figure 1]. \cite[16]{Glick19}
\end{quote}
\begin{figure}[t!]
 \centering
\includegraphics[height=6.7cm]{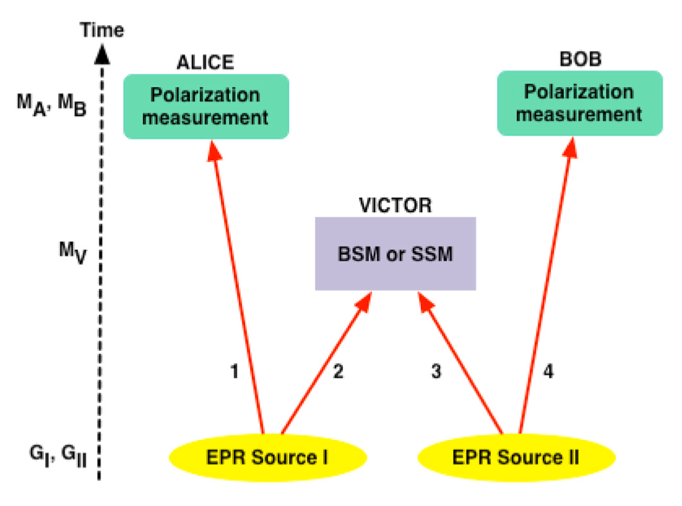}
\caption{\centering Entanglement swapping (from \cite{Glick19})}
\end{figure}

A similar procedure can be considered in which the measurement that induces the `swapping' occurs in the absolute future of the side measurements by Alice and Bob. This is called \textit{delayed-choice} entanglement swapping. As Glick describes it:
\begin{quote}
    The procedure was proposed as a thought experiment by \cite{Peres00}, but has now been realized experimentally by \cite{Ma12} and others. We begin with two entangled systems as in the ordinary case, but rather than have Victor perform his measurement prior to Alice and Bob, we delay particles 2 and 3 so that Victor can perform his measurement after his colleagues. Recall that the argument given above \ldots 
    suggests that we should expect the same result as in the ordinary swapping case. In particular, when Victor successfully performs a BSM [Bell state measurement], entanglement will be swapped to (1,4). \ldots\ [T]hese results seem to have been confirmed by an experiment conducted by \cite{Ma12} depicted  [in Figure 2]. \cite[17]{Glick19}
\end{quote}

\begin{figure}[t!]
 \centering
\includegraphics[height=6.5cm]{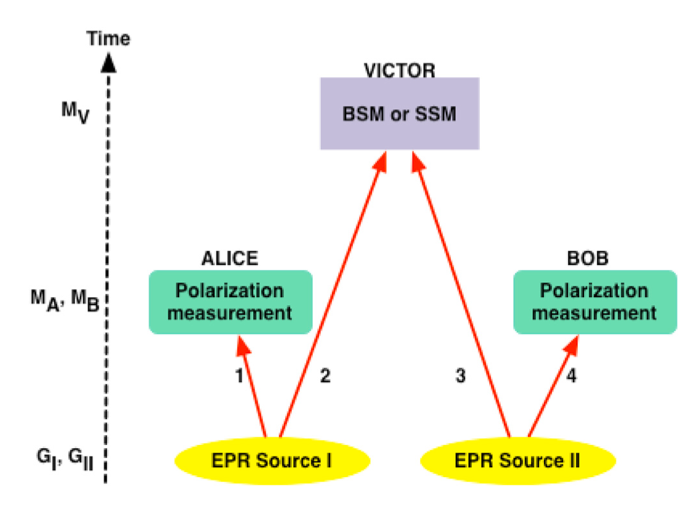}
\caption{\centering
Delayed choice entanglement swapping (\cite{Glick19}, after \cite{Ma12})}
\end{figure}

\noindent As Glick notes, this possibility seems to have peculiar consequences:
\begin{quote}
    This presents the following challenge: In the ordinary entanglement swapping case, Victor has the power to entangle (or not)
the outer particles (1,4) at a distance. In the delayed-choice case, it
seems that Victor has the same power to entangle (1,4). However, at
the time of Victor's measurement (or choice), (1,4) have already
been detected. Thus, Victor's measurement must not only be
capable of influence at a (spacelike) distance, but also backwards in
time. The only way Victor can entangle (1,4) is to act on them \textit{retrocausally} when (or before) they are measured ($t\leq M_A, M_B$). \cite[17]{Glick19}
\end{quote}

Some writers (e.g., \cite{Peres00,Healey12}) propose that we avoid this appearance of retrocausality by adopting an antirealist view of the quantum state. If the quantum state does not represent a piece or property of reality, then there is no retrocausality (or indeed any sort of causality) involved here, because there is no real \textit{effect.} Victor's measurement may change our knowledge of the past in some way, but it doesn't \textit{affect} the past.

Glick's preferences are more realist. He proposes to defend a realist view of the quantum state by allowing timelike entanglement, considering several interpretations of the resulting timelike connection.

\begin{quote}
    On this 
    view, (1,4) are entangled at $M_A, M_B$ in virtue of Victor's later measurement (at $M_V$). Generalizing, a pair of particles in a DCES experiment are entangled only if there \textit{actually is} a BSM performed in the future that swaps entanglement to them. Of course, one may not \textit{know} whether such a measurement will be performed, and hence, may wish to leave it open that the particles one encounters may be entangled. But, this doesn't trivialize entanglement as it still only applies to certain pairs of particles, namely, those prepared in an entangled state or entangled via other means (e.g., entanglement swapping). 
    \end{quote}
    
    Glick discusses two variants of this view. One is that `Victor's measurement has a retrocausal (or non-causal influence) on the pair of particles (1,4) at $t\leq M_A, M_B$.' The other, which Glick calls `Nonseparability', proposes that  `Victor's measurement gives rise to the (1,4) whole that Alice and Bob both measure.' Glick acknowledges  `difficulties in working out the details and timing of such processes', but suggests that `these difficulties are by in large the same as those already faced by [similar] approaches in the context of spacelike entanglement.' \cite[19--20]{Glick19}

Glick notes that there is an alternative explanation of the correlations involved in DCES cases, one described by \cite{Egg13}. He notes that Egg's proposal provides a halfway house between the antirealism of Peres and Healey and his own realist view, and describes it as follows:
\begin{quote}
    An alternative interpretation of DCES is given by \cite{Egg13}. Egg endeavours to provide a principled basis to accept the realist account of ordinary entanglement swapping but reject its extension to DCES. Egg's reply focuses on an aspect of Ma et al.'s DCES experiment that was omitted from the initial presentation. Unlike a simple EPR experiment, the correlations in the data recorded by Alice and Bob are only apparent once that data has been \textit{sorted} into subensembles according to the measurement performed and results obtained by Victor. Once we sort the results obtained by Alice and Bob in this way, we find that the subsets of data associated with Victor performing a BSM exhibit correlations that violate a Bell inequality. This leads Egg to conclude the following:
    \begin{quote}
     The Bell measurement on the [2,3] pair allows us to sort the [1,4] pairs into four subensembles corresponding to the four Bell states. Without delayed choice, this has physical significance, because each [1,4] pair \textit{really is} in such a state after the [2,3] measurement. But if the [1,4] measurements precede the [2,3] measurement, the [1,4] pair \textit{never is in any of these states.} This is entirely compatible with the fact that evaluating the [1,4] measurements \textit{within} a certain subensemble shows Bell-type correlations. \cite[1133, original emphasis, notation changed to match Glick]{Egg13}    
    \end{quote}
\end{quote}
As Glick says, `Egg's proposal is that we should posit physical entanglement between (1,4) only when Victor's measurement occurs before Alice's and Bob's ($M_V < M_A, M_B$).' Glick notes that this `allows one to preserve realism about entanglement (and an ontic view of the quantum state more generally) without having to adopt the revisionary metaphysics' that he himself proposes.

 Egg's analysis of the DCES cases seems to be a widespread view.\footnote{\cite{Fankhauser19} develops a similar analysis for the case of the delayed-choice quantum eraser; see also \cite{Gaasbeek10}.} The crucial point for our purposes is that if post-selection is in play, existence of Bell Inequality-violating  correlations in a subensemble of measurements need not be evidence of real entanglement. In a moment we will apply Egg's analysis to a delayed-choice version of the Delft experiment. Before that,  we need to introduce some terminology from the causal modelling literature.

\subsection{Collider bias}
In causal modelling terminology, a \textit{collider} (or \textit{inverted fork}) is a variable with more than one direct cause within a causal model. In other words, in the graphical format of directed acyclic graphs (DAGs), it is a node at which two or more arrows converge (hence the term `collider'). It is well known that conditioning on such a variable -- i.e., selecting the cases in which it takes a certain value -- may induce a correlation between its  causes, even if they are actually independent. As \cite[417]{Cole10} put it, `conditioning on the common effect imparts an association between two otherwise independent variables; we call this selection bias'.

Collider bias is sometimes called \textit{Berkson's paradox} \cite{Berkson46}. Berkson's own example involved an apparent negative dependence between diabetes and  cholecystitis in patients admitted to hospital with certain symptoms. Such symptoms tended to be caused either by diabetes or by cholecystitis, so that presence of these symptoms is a collider, in causal modelling terms. In these patients, lack of one cause does make more probable the other cause, but Berkson's point is that this is a biased sample. There need be no such correlation in the general population. 

Here's a simpler example, to lead us in the direction of quantum cases. Imagine that Alice and Bob are at spacelike separation, and play rock-paper-scissors with each other, sending their choices to a third observer, Charlie. Suppose that Alice and Bob make their choices entirely at random, and that Charlie records three kinds of outcomes: Alice wins, Bob wins, or neither wins. Obviously, post-selecting on any one of these outcomes induces a correlation between Alice's choices and Bob's choices. Equally obviously, this does not amount to real  causality between Alice and Bob.

\section{The Delft experiment and variants}

\subsection{The actual experiment}
The Delft experiment \cite{Hensen15}  
adopts a proposal originally made by Bell himself \cite{Bell79}.  Detection efficiency is improved by means of an `event-ready' measurement, whose function is to signal that two suitably entangled particles (in this case, electrons) are in the A and B detector channels. 
In the Deflt experiment, unlike in Bell's own proposal, the event-ready signal is provided by a particular outcome to a measurement that also serves to entangle the two electrons, via  entanglement swapping. The experimental procedure is described as follows:

\begin{figure}[t!]
 \centering
\includegraphics[height=10cm]{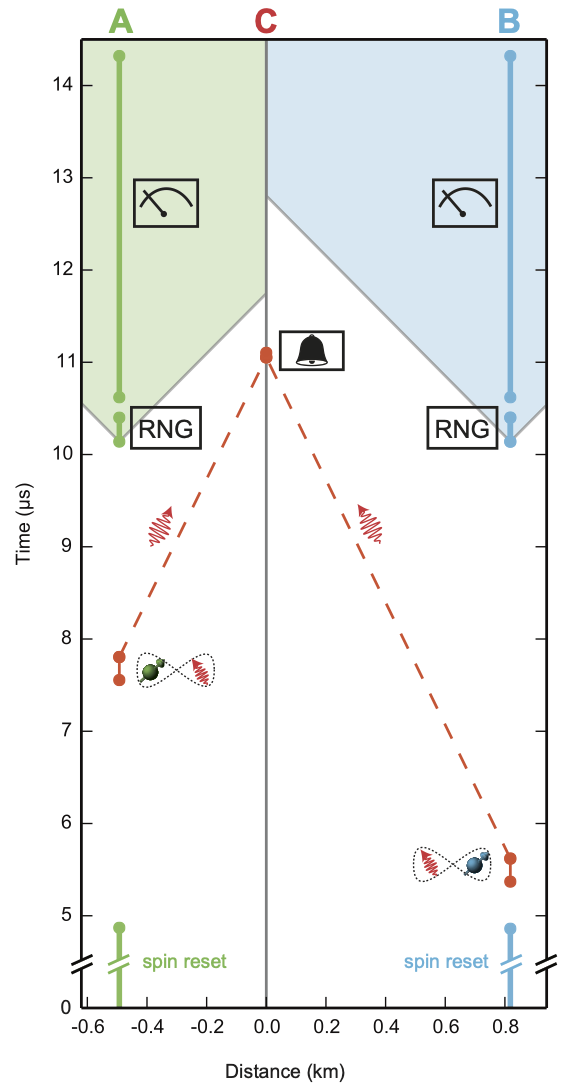}
\caption{\centering The Delft experiment (from \cite{Hensen15})}
\end{figure}

  \begin{quote}
     We generate entanglement between the two distant
spins by entanglement swapping in the Barrett-Kok
scheme using a third location C (roughly midway
between A and B \ldots ). First we entangle each
spin with the emission time of a single photon (time-bin encoding). The two photons are then sent to location C, where they are overlapped on a beam-splitter
and subsequently detected. If the photons are indistinguishable in all degrees of freedom, the observation of
one early and one late photon in different output ports projects the spins A and B into the maximally entangled state 
$\ket{\psi^{-}}=(\ket{\uparrow\downarrow} - \ket{\downarrow\uparrow})/\sqrt{2}$, where $m_s = 0 \equiv\ket{\uparrow}$, $m_s = -1 \equiv\ket{\downarrow}$. 
These detections herald the successful preparation and play the role of the event-ready signal in Bell’s proposed setup. As can be seen in the space-time diagram in [Figure 3], we ensure that this event-ready signal is space-like separated from the random input bit generation at locations A and B.  \cite{Hensen15}
    \end{quote}

In the Delft protocol, successful event-ready detection results at the central point C select a (small) subensemble of the total series of trials $E$, a subensemble we denote by $E_{C}$. These are the trials in which there is held to be successful  entanglement swapping, ensuring that the electrons at A and B are entangled. 

We note in passing that the Delft authors themselves take the view that, as they put it, `John Bell proved that no theory of nature
that obeys locality \textit{and realism} can reproduce all the predictions of quantum theory', and hence that their own result `rules
out large classes of local \textit{realist} theories' \cite[1, emphasis added]{Hensen15}. However, the claim that Bell's Theorem requires an assumption of realism is controversial; see, e.g., \cite{Norsen11} for the argument to the contrary. This is not directly relevant to our present concerns, but we mention it to emphasise that our own challenge to AAD in Delft-like experiments does not depend on this (claimed) realism loophole. 

\subsection{Delayed-choice Delftware}

Let us now consider a DCES version of the Delft experiment, Delayed Delft (DD).\footnote{Delft's famous pottery is said to have originated as a seventeenth century imitation of Chinese porcelain. It was such a good imitation that it was promptly imitated by others, producing what became known as English Delftware, Irish Delftware, etc. In the same respectful spirit we here propose a near-copy of the Delft experiment, itself inspired by the work of the Irish master.} It differs from the original experiment in that the measurement C takes place later in time, in the future light cones of the measurements A and B. This experiment would be expected to yield the same result as the original, because the relevant joint probabilities are insensitive to the relative timings of the three experiments involved. In DD, as before, let  $E_{C}$ denote the subensemble of all measurement results in which an event-ready result is recorded at C. Let $\{a_n,b_n,A_n,B_n\}$ denote the $n$th result within $E_{C}$. Here $a_n$ is the setting of measurement A,  $A_n$ the outcome of measurement A, and so on, in the usual way. 

\subsubsection{Colliders in DD}
In DD, it is uncontroversial that the measurement choices at A and B may exert a causal influence on the result of the measurement C. This is certainly so in orthodox QM, in which the measurements at A and B affect the state of the particles converging on C from the left and right, respectively. (The fact that the measurements at A and B take place \textit{before} that at C is crucial in this account, of course.) 

In causal modelling terms, then, the outcome of the measurement C is a collider, for causal influences originating at A and B. 
This immediately  puts on the table the possibility that any apparent AAD between A and B might be a manifestation of selection bias. We say `puts on the table' here because in principle an association between A and B might result from a \textit{combination} of selection bias at C and some real underlying causal influence between A and B.

The possibility that the apparent AAD between A and B might be a selection artifact is our \textit{Collider Loophole} (CL). But it is one thing to identify a possibility, another to show that it is actually the case. How can we determine whether CL is the true explanation of the apparent AAD across DD? We will proceed by offering two tests for genuine causality, and explaining why DD seems to fail both tests. 

\subsubsection{The No Difference test}

The first test starts with this question. Would  $A_n$ or $B_n$ have been different, if the measurement C had not (later) taken place?\footnote{As a further refinement to the experiment, we might imagine that whether the measurement C takes place is itself determined by a randomiser, also in the absolute future with respect to A and B.} 
The intuitive answer to this question is `No'. Because C lies in the future with respect to the measurements A and B, allowing it to influence the measurement results at A and B would amount to retrocausality, which we are assuming is impossible (NoRA). 

Yet answering `No' leads to a puzzle. If each individual measurement at C makes no difference in this way, then it seems to follow that the entire set of C measurements makes no difference -- in other words, that \textit{all} the measurement results in  $E_{C}$ would have been the same, even if the measurement device at C had simply been absent altogether. But these results display Bell correlations between A and B, and hence seem to constitute evidence of entanglement between the electrons A and B. How could such entanglement arise, if there is no measurement to provide entanglement swapping?
Egg's approach to DCES answers this question, telling us that the apparent entanglement is a selection artifact. The set of results $E_{C}$ is indeed just the same, whether or not C measurements take place, but there's no real entanglement in either case.  

It is easy to see how this reasoning extends to AAD. If the results in $E_{C}$ are independent of whether the C measurements actually take place, then the questions of AAD in the two cases -- i.e., with and without the C measurements -- stand and fall together. Either we are committed to AAD even in the absence of the C measurements, or we are not committed to it in the presence of the C measurements. And the latter is by far the more plausible option. Like entanglement itself, on the Egg view, the appearance of AAD is a selection artifact. 

What we have just done is to take reasoning that supports the Egg view -- the argument that in the absence of retrocausality, C can {make no real physical difference} in the DECS case -- and applied to the issue of AAD. Let's call this the \textit{No Difference Argument} (NDA) against AAD in DD. This is our first causal test, and we have argued that DD fails it. 
To get to our second causality test we'll proceed indirectly, via a possible objection to NDA.

\subsubsection{Thinking about counterfactuals}
NDA appealed to the assumption that  the results in $E_{C}$ are independent of whether the future C measurements take place -- that C makes no difference. A possible reply is that even if C doesn't make a difference to the \textit{actual} contents of $E_{C}$, it might make a difference to the kind of \textit{counterfactuals} that would support a claim of causal influence from A to B. If so, that might explain how a relation of causal dependence could exist in the presence of C that would not exist in its absence.  

To explore this proposal, let's think of an example. Let's consider simply the extreme correlations, the ones that are relevant in the original EPR argument. Suppose that Alice chooses setting $a_n=0$, and is told that the case falls in $E_{C}$.\footnote{In case this supposition should seem controversial, we note that the argument in this paragraph could be phrased entirely from the point of view of the experimenter at C, so that it is this experimenter (hereafter `Charlie') who raises the question whether the case would have still be in $E_{C}$, if Alice had chosen differently. We phrase it in Alice's voice to stay as close as possible to the reasoning in the original EPR argument.} From the fact that $E_{C}$ satisfies the Bell correlations, it follows that if Bob has chosen the same setting $b_n=0$, then $A_n=-B_n$. In other words, in the spirit of the EPR argument, Alice knows something about the probabilities on Bob's side of the experiment. If $A=1$, for example, then she knows that $Pr(B = -1\vert{b_n=0})=1$. Now the crucial question. Does she also know that \textit{had} she instead chosen $a_n=1$, she would have been able to predict the result of a measurement with $b_n=1$ with similar certainty? No -- for she doesn't know that the run of the experiment would have fallen within $E_{C}$, in that counterfactual case. If we \textit{assume} that it does so -- if we hold fixed the result of the C measurement, in effect -- then the reasoning goes through, but why should we be entitled to do that? Why should C not be sensitive to the choice of measurement setting at A, in a way that makes it possible that if Alice had chosen differently, the result of the C measurement might have been different? This thought will lead us in the direction of our second causality test.

\subsubsection{The Counterfactual Fragility test}
NDA offered one reason for thinking that violation of Bell inequalities in DD should be seen as an artifact of post-selection, rather than a manifestation of AAD. There is a second factor that points in the same direction. Any argument from a set of experimental data to causal dependence is going to require an assumption something like the following.  
\begin{quote}
 \textit{Alternative Measurements (AM)}---It is legitimate to consider measurements
which the experimenters might have performed, in addition to those which
they actually perform. \cite{Clifton90}   
\end{quote}
In this case, the assumption AM is formulated in a discussion of Bell-style arguments, but the point is much more general. In the causal modelling framework, it is embodied in the assumption that exogenous variables may take a range of values.
If an ensemble of correlation data is to provide information about causation, it needs to respect such a principle. It needs to provide information  about the results of alternative choices. But a post-selected ensemble may fail to do so. It may be `counterfactually fragile', in the sense that it doesn't support inferences about what would have happened, had an alternative measurement been performed. To support a causal claim, a set of data needs to be counterfactually \textit{robust.} 

Here's a simple example. Suppose I record occasions on which it is true either that I wear green socks in the morning and it is sunny the same afternoon, or that I wear red socks in the morning and it is raining in the afternoon. `Look', I claim, `I can control the weather, at least in this subensemble of cases.' What I've missed is that the selection method for the subensemble doesn't respect AM, and is hence counterfactually fragile.  Had I chosen the other sock colour the resulting case would not have been in the subensemble at all, in most cases. 

It is easy to see how this kind of fragility might be present in the DD protocol. As we just observed, there is no guarantee that if A had chosen an alternative measurement setting in case $\{a_n,b_n,A_n,B_n\}$, the resulting measurement would have been in the subensemble $E_{C}$. Why not? Because there is no guarantee that the result of the measurement C would have been the same in that case. As we noted, conventional QM takes it for granted that setting choices can influence each of the particles converging on the central vertex of the \W\, and in turn influence the outcome of the measurement at C.

The upshot is that the correlations \textit{within} the subensemble $E_{C}$ may provide no guide to what $A_n$ and/or $B_n$ \textit{would have been,} had $a_n$ been different. If so, then the data provided by $E_{C}$ is not counterfactually robust, in the sense described above, and cannot support the causal claim of AAD.  Let's call this the \textit{Counterfactual Fragility Argument} (CFA).  It is our second causal test, and again, there are good grounds to think that DD fails it.

\subsubsection{Summary: the Collider Loophole}

We have argued that the case for cross-\W\ AAD in DD is vulnerable to the Collider Loophole (CL). In other words, Bell correlations revealed in the results of DD are likely to be an artifact of collider bias, rather than a sign of genuine AAD. It is uncontroversial that C is a collider in DD, and the diagnosis of collider bias is supported by intuitive tests for causal dependence in two ways. 

First, from the fact that C occurs later than A and B, and NoRA, we concluded that the results in the subensemble $E_{C}$ would have been the same, even if the C measurements had not taken place. NDA therefore undermines the claim that the correlations in $E_{C}$ reveal genuine AAD. 

Second, the fact that C occurs later than A and B allows the measurement settings at A and B  to influence the result of the measurement at C. But this means that $E_{C}$ may be counterfactually fragile. So CFA, too, undermines the claim that the Bell correlations in DD reflect any genuine causal dependency.

All of these points appealed to the fact that in DD, C lies in the absolute future of A and B. This suggests that  CL would be avoided by a version of the Delft experiment that put C in the past with respect to A and B. We turn to that case in \S5, before returning to the actual Delft experiment (which lies between these two variants, in the sense that it puts C at spacelike separation to A and B).

We emphasise again that CL does not challenge 
AAD  \textit{within each of the two wings} of DD. In each of the wings we have a component with \V\ geometry, within which there is no further collider to generate selection bias.\footnote{For clarity, we note that there is a different post-selection loophole that is relevant within the \V\ geometry. As \cite{Blasiak20} explain, Bell experiments with this geometry always rely on some sort of post-selection -- e.g., the fact that two particles were both detected in the same time bin -- in generating their experimental data.  These authors discuss the possibility that this kind of post-selection creates experimental artifacts, and propose a criterion to eliminate this possibility. It requires that the post-selection still be possible if we drop the data from any single post-selection measurement.  But the post-selection at the intermediate vertex in the \W\ geometry of the Delft experiment cannot be eliminated.\label{footnote_1}}  
 It may be that such ‘mini-AAD’ has a crucial role in the processes that make post-selection possible in the \W-geometry as a whole, enabling the C measurement to ‘know about’ the A and B measurements. Our point is that this need not imply AAD across the \W\ from A to B or vice versa. The Collider Loophole stands in the way.

\section{Connections to Bell's Theorem}

Before we turn to other variants of the Delft experiment, we want to clarify the relationship between the above discussion and Bell's Theorem. As we said in \S1, we have been using the term AAD deliberately, avoiding the term 'nonlocality'. We are now in a position to explain this choice.

Bell derived a family of inequalities from two mathematical assumptions, widely known as \textit{Local Causality} (LC) and \textit{Statistical Independence} (SI).  When these inequalities are violated, as they are for the Bell correlations observed in the entanglement experiments we are discussing, at least one of these assumptions must fail.  It is instructive to ask how the Collider Loophole might accomplish this, if there is to be no AAD across the \W\ as a whole.

LC is formalized by Bell himself \cite{Bell90, Norsen11} as a conditional independence (or screening) condition between two spacelike-separated wings of an experiment.  
Norsen describes Bell's own account of LC with reference to the diagram reproduced here as Figure 4, adapted from \cite{Bell90}.
\begin{figure}[t!]
 \centering
\includegraphics[height=4.5cm]{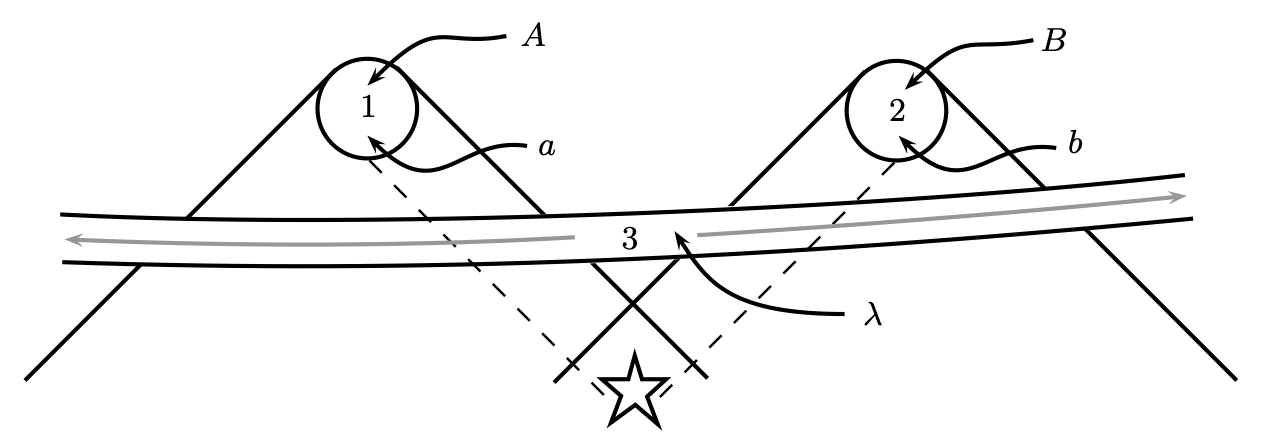}
\caption{\centering Bell's notion of Local Causality (from \cite{Norsen11})}
\end{figure}
Norsen's caption for this diagram reads as follows:
\begin{quote}
Spacetime diagram illustrating the various beables of relevance for the EPR-Bell setup. \ldots\ Separated observers Alice (in region 1) and Bob (in region 2) make spin-component measurements (using apparatus settings ${a}$ and ${b}$ respectively) on a pair of spin- or
polarization-entangled particles (represented by the dashed
lines). The measurements have outcomes $A$ and $B$ respectively. The state of the particle pair in region 3 is denoted
$\lambda$. Note that what we are here calling region 3 extends across
the past light cones of both regions 1 and 2. It thus not only
``completely shields off from 1 the overlap of the backward
light cones of 1 and 2'', but also vice versa. Bell’s local causality condition therefore requires both that ${b}$ and $B$ are irrelevant for predictions about the outcome $A$, and that ${a}$ and $A$ are irrelevant for predictions about the outcome $B$, once $\lambda$ is specified.
\end{quote}

As Norsen goes on to say:
\begin{quote}
A complete specification of beables in this region 3 will therefore, according to Bell’s concept of local causality, ``make events in 2 irrelevant for predictions about 1'' and will also make events in 1 irrelevant for predictions about 2.
\end{quote}
In formal terms, LC may thus be written:
\begin{eqnarray}
\label{eq:LC}
P({A}|a,b,{B},\lambda)&=&P(A|a,\lambda),\\
P({B}|a,b,{A},\lambda)&=&P(B|b,\lambda). \nonumber
\end{eqnarray}
In the same terminology, Statistical Independence (SI) between the settings ($a,b$) and the prior state of the system ($\lambda$) is the following condition: 
\begin{eqnarray}
\label{eq:SI}
P(\lambda |a,b)=P(\lambda) 
\end{eqnarray}
Norsen emphasises the character of the probabilities that Bell has in mind in his definition of LC.
\begin{quote}
    Bell has deliberately and carefully formulated a local causality criterion that \ldots\ is \ldots\ stated explicitly in terms of probabilities -- \textit{the fundamental, dynamical probabilities assigned by stochastic theories to particular happenings in space-time.} Note in particular that the probabilities in Eq.~(1) are not subjective (in the sense of denoting the degree of someone’s belief in a proposition \ldots), they cannot be understood as reflecting partial ignorance about relevant beables in region 3, and \textit{they do not (primarily) represent empirical frequencies for the appearance of certain values} \ldots. They are, rather, the fundamental ``output'' of some candidate (stochastic) physical theory. \cite[10, emphasis added]{Norsen11}
\end{quote}
We call attention to Norsen's distinction between two notions of probability because it turns out to be a helpful way to make the point that when post-selection is in play, LC and SI may each be applied in two different ways, \textit{within the same model.} 

Let's begin with a toy example.  With reference to Figure 4, imagine that Alice and Bob (in regions 1 and 2, respectively) each generate ordered pairs of (genuinely) random bits and send them to a third observer, Charlie, in their absolute future. Think of the first of each pair of bits as a `setting' and the second as an `outcome'. Let Charlie perform a binary measurement on each pair of ordered pairs of bits, the producing positive results with probabilities based on the Bell correlations. Selecting for positive outcomes at C will then generate an ensemble of results of the form $\{a_n,b_n,A_n,B_n\}$. Denote this ensemble $E_{C}$, as before.

The selection procedure guarantees that the ensemble $E_{C}$ violates a Bell inequality. Bell's Theorem therefore implies that either LC or SI must fail, within this ensemble of results. But SI is guaranteed by the assumption that the settings are genuinely random bits, 
 so the effect of post-selection must be to induce an LC-violating correlation between the the two sides of the experiment, conditional on  $\lambda$.
In other words, at least one of the following must hold:
\begin{eqnarray}
\label{eq:LCps}
P_{ps}({A}|a,b,\lambda) \neq P_{ps}({A}|a,\lambda) \\
P_{ps}({B}|a,b,\lambda) \neq P_{ps}({B}|b,\lambda) \nonumber
\end{eqnarray}
 where $P_{ps}$ denotes probability (i.e, in this case, frequency) within the post-selected ensemble. 

Let us call this result a violation of LC$_{ps}$, adding the subscript to remind ourselves where the probabilities involved originate.
Does this violation of LC$_{ps}$ reflect any real causality between A and B? Clearly not. It is simply an artifact of collider bias. Alice's and Bob's pairs of bits are joint causes of the outcome of Charlie's measurement, and that's the source of the correlation.  Equally clearly, the violation of LC$_{ps}$ does not imply any violation of LC as interpreted in terms of the underlying dynamical probabilities of the model. (Let us write LC$_{dp}$ for this case.) In this toy example, these fundamental probabilities are those involved in the stipulation that Alice and Bob generate genuinely random bits, so that LC$_{dp}$ is trivially satisfied:
\begin{eqnarray}
\label{eq:LCdp}
P_{dp}({A}|a,b,{B},\lambda) = P_{dp}(A|a,\lambda) = 0.5,\\
P_{dp}({B}|a,b,{A},\lambda) = P_{dp}(B|b,\lambda) = 0.5. \nonumber
\end{eqnarray}

The need to distinguish LC$_{dp}$ and LC$_{ps}$ explains our caution about using the term `nonlocality'. This simple example shows that in the presence of post-selection, we may have nonlocality in the sense of failure of  LC$_{ps}$  (call this `nonlocality$_{ps}$'), without nonlocality in the sense of failure of LC$_{dp}$. As the example illustrates, nonlocality$_{ps}$ does not imply AAD, or genuine causality.\footnote{Astute readers will have noticed that we have exaggerated the difference between the two kinds of probability $P_{ps}$ and $P_{dp}$, in order to make the distinction between LC$_{ps}$  and LC$_{dp}$ more striking. $P_{ps}$ could also be explained in terms of the underlying dynamical probabilities, $P_{ps}$ simply resulting from $P_{dp}$ by conditioning on an outcome of the measurement at C.}  

This lesson carries over to DD. The diagnosis of the Bell correlations offered by the Collider Loophole excludes cross-\W\ AAD, but it is not incompatible with nonlocality$_{ps}$. DD may well exhibit nonlocality$_{ps}$, despite respecting LC$_{dp}$ itself. 
In other words, the distinction between LC$_{dp}$ and LC$_{ps}$ explains how it can be true both (i) that a failure of LC explains the Bell correlations observed in DD, and (ii) that the experiment exhibits no cross-\W\ nonlocal causation. The apparent tension between these claims dissolves when we realise that they rely on different applications of LC, (i) relying on the LC$_{ps}$ sense and (ii) relying  on the LC$_{dp}$ sense.

Now to SI. Failure of Bell inequalities in DD requires that one of LC and SI fails, and we have been considering the possibility that it is LC, in the form LC$_{ps}$. However, SI might also fail in a post-selected ensemble. After all, the geometry of DD allows the central measurement to be informed of the remote measurement settings at A and B via a classical channel. This post-selection might induce a correlation between $a$ and $b$ and $\lambda$, violating SI$_{ps}$ (where, again, the subscript makes explicit that we are referring to the post-selected ensemble). Again, violation of SI$_{ps}$ need not imply a violation of SI$_{dp}$.

Indeed, a small modification of the toy example above illustrates this possibility. Suppose now that the random bits interpreted as outcomes in that example are generated as a pair ${A,B}$ at the source marked with a star in Figure 4, and this pair is sent directly to Charlie. Alice and Bob each generate a random `setting' bit, $a$ and $b$, respectively, and send these to Charlie. Charlie post-selects, as before, to yield a subensemble $E_{C}$ violating a Bell inequality. Once again, this implies correlations between $a$ and $B$ and/or between $b$ and $A$, within the post-selected ensemble. Because the pair ${A,B}$ now falls within $\lambda$, however, these correlations manifest as a failure of SI$_{ps}$, not as a failure of LC$_{ps}$.

These toy examples suggest that in any real experiment in which apparent AAD is an artifact of collider bias, the precise explanation in terms of LC$_{ps}$ and SI$_{ps}$ is likely to be model-dependent. For this reason we will not discuss these issues further in this piece. We will continue to speak of AAD, rather than `nonlocality', to avoid the need to keep in mind the distinction between LC$_{dp}$ and LC$_{ps}$.

  \begin{figure}[t!]
  \centering
\includegraphics[height=6.5cm]{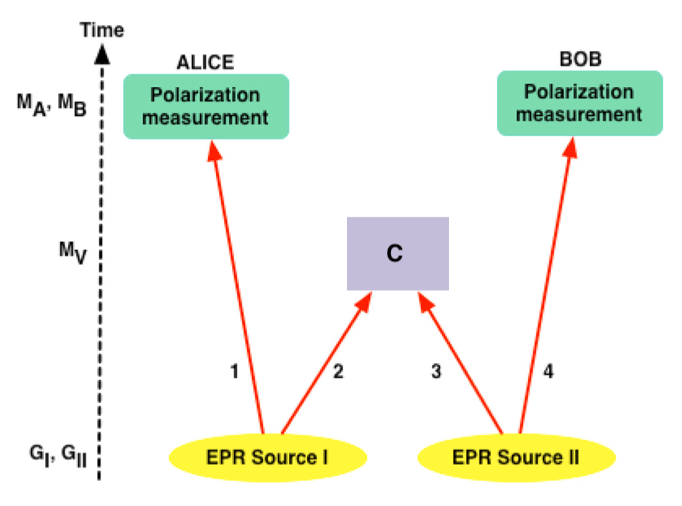}
\caption{\centering\small Early Delftware (entanglement swapping in the past)}
\end{figure}

\section{Early Delftware}
We now turn to the case in which the measurement C is in the absolute past of A and B (Figure 5), a case we'll call Early Delft (ED).\footnote{This is the geometry of three loophole-free experiments that followed the original result from Delft \cite{Giustina15,Shalm15,Rosenfeld17}.} In this case, the geometry seems to do a much better job of avoiding  CL. The assumption NoRA ensures that A and B are not causes of C, at least partially removing the threat that C is a collider, and hence a source of collider bias. (We say `partially' because C is still a collider with respect to influences from the two sources. More on this below.) 

More importantly, our two causal tests  no longer seem a threat. In DD, the first test, NDA, relied on the claim that all the individual results in $E_{C}$ would have been the same, even if the C measurements had not been made. But that claim relied on NoRA; and NoRA cuts no ice in ED, obviously, where C is earlier than A and B. 

As for the second test, CFA, it relied in DD on the possibility that the choice of measurement settings at A and B could influence C, at least in principle. As a result, we were not entitled to infer that a given result would still have been in $E_{C}$, even if Alice had chosen a different measurement. That was why $E_{C}$ was susceptible to counterfactual fragility. But now, given NoRA, 
 the measurement choices at A and B cannot influence C. CFA is blocked.

\subsection{Determined scepticism?}

\begin{figure}[t!]
 \centering
\includegraphics[height=5cm]{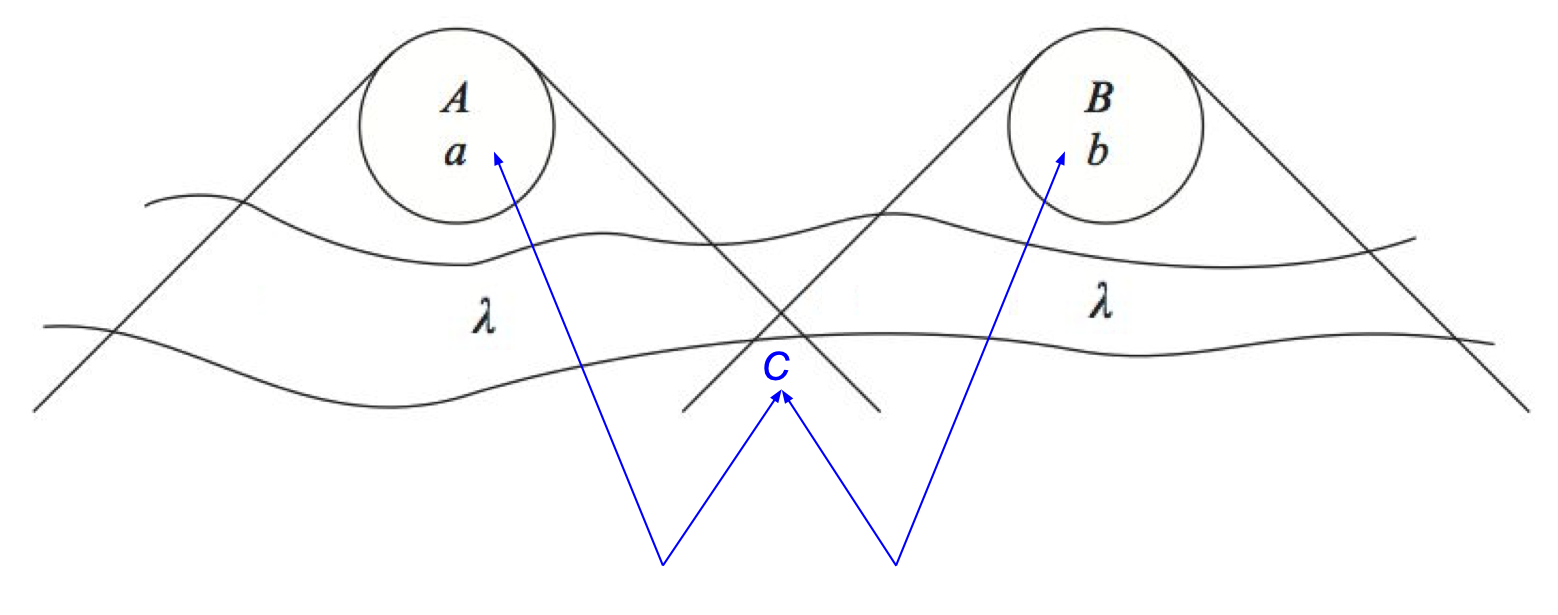}
\caption{\centering Early Delft in Bell's framework (adapted from a diagram in \cite{Bell04})}
\end{figure}

At this point, we introduce a character we'll call the \textit{AAD Sceptic,} whose mission in life is to find small loopholes in arguments for AAD. Concerning ED, the AAD Sceptic objects that although passing the two causal tests blocks two arguments \textit{against} AAD in ED, it doesn't actually confirm that there \textit{is} AAD in ED. After all (the Sceptic continues),  there's still a collider of some kind in ED – the central measurement at C is presumably a joint effect of the events that supply the particles from left and right. As in DD, ED will only yield Bell correlations when we condition on the result of the measurement at C. Moreover, the correlations as a whole in ED are exactly
the same as in DD, where we've agreed that there's a strong case for saying that there
isn't AAD. In the light of this,  isn't it weird to think that a small difference in
C's location could make such a big difference to the causal structure of the case?

How should we meet this argument? One strategy would be to appeal to Bell's
Theorem, where ED permits a straightforward application of Bell's own reasoning.  Consider Figure 6, where the measurement C occurs in the past lightcones of A and B.  Unlike in DD, this measurement can be considered part of the preparation of the final two-particle system, contributing to correlations in  $\lambda$.  The violation of Bell inequalities observed between A and B therefore implies that one of the assumptions LC and SI must fail, in the corresponding ensemble of results. 

In the light of our discussion in \S4, however, it is clear that this won't be enough to convince the AAD Sceptic. The Sceptic will repeat that even in ED, Bell correlations are only revealed in subensembles, corresponding to particular results of the measurement at C. So ED still gives us only a violation of LC$_{ps}$, not a violation of LC$_{dp}$. There's still a collider of some kind at C, so why should a violation of LC$_{ps}$ be enough to guarantee AAD? It wasn't in DD.

To meet this challenge, we need to identify some difference between the nature of the colliders in DD and ED -- something that will explain how Bell correlations can be a selection artifact in DD but not in ED. The crucial point seems to be the difference in the causal relationship between the measurement settings at A and B and the outcome at C, in the two cases. In DD, as we've seen, it is uncontroversial that the measurement settings at A and B may themselves be contributing causes to the outcome at C. This means that the settings themselves `feed into' the collider at C. In a conventional QM picture, this happens because the measurements at A and B affect the state of the other particle in the corresponding pair, before those particles arrive at C. Intuitively, we might say that this feeds information about the settings at A and B in the direction of C.

In ED, however, the conventional picture has information, or causal influence, flowing the other way -- \textit{from} the measurement C \textit{to} the particles due to be measured at A and B.  Intuitively, this makes C a simpler kind of collider. Without input from the measurement settings, it simply isn't rich enough to generate Bell correlations by post-selection. 

We have some sympathy with the AAD Sceptic in thinking that it's counterintuitive that a small difference in the location of C could make such a big difference to the causal structure of a \W-geometry Bell experiment. However, the difference between ED and DD certainly isn't negligible. Sitting between them is the third case,  in which C is spacelike separated from A and B. If there's any merit in the Sceptic's point, it might be expected to emerge from the three-way comparison between ED, DD, and this intermediate case. This brings us nicely to the actual Delft experiment, which has this intermediate geometry.

\section{The Collider Loophole in the real Delft experiment}

Recapping the argument so far, we have considered two variants of the Delft experiment. We argued that in one (DD) but not apparently the other (ED), CL would be a significant challenge to the claim that its results provided evidence of AAD, across the \W\ geometry concerned. With these two variants of the Delft experiment as comparison cases, we now return to the actual Delft experiment. 

As just noted, the actual Delft experiment is an intermediate case between ED and DD, in the sense that the C measurements are at a spacelike separation to the corresponding A and B measurements. As \cite[3]{Hensen15} say, they `ensure that [the] event-ready signal is space-like separated from the random input bit generation at locations A and B'. Where does this leave it? Is it troubled by the Collider Loophole, like DD, or safe from it, like ED?

 {As we shall see, this question turns out to be quite subtle. It is also in one sense moot, because we now have three similar Bell tests with the ED geometry \cite{Giustina15,Shalm15,Rosenfeld17}. But the Delft experiment was the first Bell test to use this technique, and so for historical as well as metaphysical reasons, it is interesting to ask on which side of the line it falls. Was there really AAD `across the \W' in Delft in 2015?} 

Let's see how far we can push the case for the Collider Loophole in this case. Could the correlations in $E_{C}$ be merely a selection artifact, as in DD? Can we still use NDA to argue that they would have been the same, even if C had not taken place? In DD we appealed to NoRA to argue that the C measurements makes no difference to the A and B measurements. 
As a substitute we might now try to appeal to a principle of `no spacelike causality', calling on the  authority of special relativity in a familiar way. 

There's a very obvious objection to such an appeal in this context, but set that aside for the moment. If we could appeal to such a principle, it would again tell us that the A and B measurements would have been the same without the C measurements, and hence that the actual subensemble  $E_{C}$ would have involved the same correlations, even without C. As before, no one would claim that there would be genuine AAD from A to B without C; so this would be enough to cast doubt on the claim that the correlations in  $E_{C}$ reveal genuine AAD, even in the case in which we do have the C measurements.

Summing up, a no spacelike causality principle would leave the Delft experiment vulnerable to NDA. On the other hand, it would provide some protection against CFA. Without spacelike causality, the A and B measurement choices cannot affect the result of the C measurement, which was the main source of concern about counterfactual fragility. So the news for the Delft experiment would be mixed: continuing vulnerability to NDA, but protection against CFA. CL would still be a concern.\footnote{At this point we might improve the news by a move we proposed in \S3.2.3 -- more on this in \S6.1.2 below.\label{footnote_3}}

The obvious objection set aside a moment ago is that, as Bell himself has shown us, QM implies that there \textit{is} spacelike causality. So we can't glibly assume its absence, in this context. 
Unfortunately for the project of extracting a watertight case for AAD from the Delft experiment, this point cuts both ways.   If we can't exclude the possibility that the C measurement makes a difference to the A and B measurements, we can't exclude the possibility that the A and B measurement settings make a difference to the result of the C measurement. And that leads us vulnerable to counterfactual fragility. We can't assume that if $a_n$ had been different, the instance in question would still have fallen within  $E_{C}$. 

\subsection{A preferred frame to the rescue?}
Does it help the case for AAD if we allow a preferred frame, and then revert to our previous assumption -- no retrocausality, in the sense of the preferred frame?\footnote{Without a preferred frame, NoRA itself excludes spacelike causality, a point noted by Einstein as early as 1907.  \cite[235]{Bell04} cites a discussion by Einstein in which Einstein observes that spacelike causality would imply retrocausality in some frames.}
This does indeed help if C is earlier than A and B in the preferred frame. Then C may make a difference to A and B, so that we can't assume that   $E_{C}$ would have been just the same, without the C measurements. And yet A and B can't make a difference to C, weakening the case for an appeal to counterfactual fragility. With both causal tests blocked, the Collider Loophole seems sealed, just as in ED.\footnote{The assumption of no retrocausality with respect to the preferred frame will also do the work that NoRA does in blocking the AAD Sceptic in \S5.1} 

This approach suffers from the disadvantage of needing a preferred frame, but that's a cost that many in the field are in any case reconciled to paying.  
Setting that cost aside, does it provide the Delft experiment with a complete solution to  the challenges of CL? We want to identify two arguments for pessimism. The first is our reason for concluding that the Delft experiment cannot entirely evade CL. The outcome of the second argument is less clear. We think it raises issues that deserve to be put on the table in this context, but we also propose a way to escape it.

\subsubsection{Which is the preferred frame?}
The first argument turns on the observation that it is going to be difficult, probably impossible, to exclude the possibility that in the actual experiment, C is in the future of A and B with respect to the preferred frame. That option inherits the problems of DD, not the safety of ED. Who is to say which of all the possible preferred frames is the `true' preferred frame? Without an answer to that question, the Delft experiment can't exclude the possibility that its actual results are a selection artifact.\footnote{When we introduced CFA in \S3.2.4, it was important that C was in the future lightcone of A and B.  This would not be the case here, but the present  line of thought assumes that only time-order, not lightcones, determine allowed causal influences.  So the earlier arguments would still apply, with `future lightcone' simply replaced by `future relative to the preferred frame'.} 

In case the reader feels tempted to wave this concern aside, it is worth emphasising that the Delft experiment belongs to a decades-long project of closing what many have seen as tiny loopholes in the experimental case for quantum AAD. It would hardly be in the spirit of that project simply to wave aside this new concern, arising from the experimental undetectability of the assumed preferred frame.

\subsubsection{Does the argument for AAD beg the question?}
The second argument turns on the question whether an argument for AAD based on the Delft experiment begs the question, even granting a preferred frame. To avoid NDA  -- i.e., the challenge that  $E_{C}$ would have been just the same without the C measurements -- it was necessary to allow that the C measurements might make a difference to the A and B measurements. But in the spacelike case -- i.e., in the actual Delft experiment  -- such an influence of C on A and B would itself be a case of AAD. This leaves the argument in a delicate position. Unless it already \textit{assumes} AAD from C to A and C to B, it is unable to meet a challenge to its claim to establish AAD between A and B. This looks dangerously like a logical circle. 

It might be replied that we already have evidence for AAD in the two component wings of the experiment (from C to A, and from C to B), 
  and that we are entitled to rely on this evidence to block a challenge to the claim that the experiment demonstrates AAD from A to B. But if that's the way the logic works, it deserves to be made explicit. Again, there's a question as to whether it is good enough, by the lights of the project of making the case for AAD completely watertight. Imagine again our  AAD Sceptic, keen to exploit any possible loophole to try to refute AAD. Such an opponent is hardly going to be convinced by a proposed reason for setting aside CL, if that proposal assumes AAD somewhere else.

In our view, a better reply is to appeal to an suggestion we considered in \S3.2.3. There, we proposed that even if C doesn't make a difference to the \textit{actual} contents of $E_{C}$, it might make a difference to the kind of \textit{counterfactuals} needed to support a claim of causal influence from A to B. In the context of DD, this proposal didn't seem to work. On the contrary, the fact that A and B might influence C gave us a reason to think that the counterfactuals needed would not hold -- that led us to CFA.

In the present context, however, a ban on AAD would prevent the A and B measurement settings from affecting C, saving the counterfactuals. This gives the Delft experiment an answer to the AAD Sceptic. If the Sceptic were right, we would be able to appeal to this different way of meeting the causality tests. This seems to avoid the charge that the inference from the Delft results to AAD is begging the question, though it is a subtle matter, and we think that the reasoning needs to be made clear.    
Let's lay out the reasoning explicitly:
\begin{enumerate}
    \item Assume for the sake of the argument that the AAD Sceptic is right, and hence that there is no AAD anywhere in the Delft experiment (either in the wings or across the \W).
    \item Then the move proposed in \S3.2.3 (see also fn.~\ref{footnote_3}) protects the Delft argument from NDA and CFA.  With these causality tests met, the observed violation of the Bell inequality shows that the AAD Sceptic and  Assumption 1 are wrong.
    \item This shows that there is AAD \textit{somewhere,} but this isn't a complete `bridge repair' -- i.e., a complete defence of AAD across the \W\ -- because the AAD might be only in the wings. However, it does protect the Delft argument against the charge that that it is begging the question.
\end{enumerate}

\subsection{Summary: CL in the actual Delft experiment}

Summarising our discussion, the best prospect for defending the Delft experiment against CL lies in the assumption of a preferred frame, and reliance on NoRA with respect to this frame. So long as C is earlier than A and B with respect to the preferred frame, the argument for AAD in the Delft experiment escapes CL, and avoids passes the causal tests NDA and CF, in the same way as in the case of ED. 

However, the cost of this move --  setting aside the theoretical cost of reliance on a preferred frame -- is that safety from CL becomes experimentally unconfirmable. In any actual version of the Delft experiment, it would simply be unknown whether the required condition was satisfied.

In addition -- echoing the AAD Sceptic's point in \S5.1 about the difference between ED and DD -- we note that reliance on a preferred frame commits us to a sudden and unobservable change in the causal structure of the experiment, as the spatiotemporal location of C varies with respect to that of A and B. Again, many in the field may feel that this is not much of a cost, because they are committed to such things for other reasons. (Think of the sensitivity of the causal structure of a regular \V-shaped EPR-Bell experiment to the time-order of measurements, in any collapse model.) But it is worth asking whether there are models that avoid this consequence, especially given that the experimental correlations are independent of the temporal location of C. We discuss this issue in Appendix A.

\section{Conclusion}
The use of entanglement swapping in Bell tests introduces an additional loophole into arguments from violation of the Bell inequalities to AAD.  Under conventional assumptions -- i.e., excluding retrocausality -- the sensitivity of such experiments to this Collider Loophole depends on the temporal relation between the entanglement-swapping measurement C and the measurements A and B. CL a threat if the C is in the future of A and B, but not if it is in the past. The Delft experiment is the intermediate case, in which the separation is spacelike. We  argued that this leaves it vulnerable to CL, unable to confirm experimentally that it avoids it.\footnote{We are very grateful for comments from Gerard Milburn, Peter Evans, David Glick, Travis Norsen, Malcolm Perry, and two anonymous journal referees.}

\begin{appendices}

\renewcommand\thesubsection{\thesection.\arabic{subsection}}
\renewcommand\thefigure{\arabic{figure}}
\renewcommand\thetable{\arabic{table}}
\newpage
\section{Does retrocausality make a difference?}
In the body of the paper we adopted the conventional assumption that there is no retrocausality. NoRA  played a crucial role in our argument, at several points. 
What difference would  relaxing this assumption -- i.e., allowing retrocausality -- make to the conclusions reached above? We want to make three main observations, beginning with some clarificatory remarks about the retrocausal option in general. 

\subsection{The retrocausality `loophole' in general}
   
Retrocausal models reject a key assumption in the derivation of the Bell inequalities, and hence claim to offer a further loophole -- in this case logical, not experimental -- in the case for AAD in the quantum world. To be precise, they reject Statistical Independence (SI), the assumption that any hidden variables are independent of future measurement settings. However, their route to rejection of this assumption is not the one familiar in much of the literature concerning Bell tests. The familiar  route to the rejection of SI involves the hypothesis that measurement settings and hidden variables might have some common cause in their joint past, a suggestion which is (rightly) seen as in tension with the assumption that measurements settings can be treated as exogenous variables. This is the so-called `free will' issue. The retrocausal proposal, involving a direct causal influence from the future settings to the past hidden variables, is claimed to involve no difficulties of this kind. It treats measurement settings as exogenous variables in the normal way, but postulates that some of the effects of these variables lie in the past;  see \cite{
FriedrichEvans19, WhartonArgaman20, NorsenPrice21} for recent discussions.

The retrocausal option thus presents a very general challenge to the argument from Bell correlations to AAD, a challenge applicable in the case of \V\ geometries. However, as we noted in \S1, retrocausal models do allow causality `across the \V' in Bell experiments. Where they differ from models admitting AAD is in taking this causal influence to be indirect -- a product of a zig zag causal path, via the two arms of the \V.  

\subsection{Retrocausality and entanglement swapping}
This brings us to our second observation, about the effect of admitting this retrocausal possibility in the \W\ cases. We want to make two comments. First, this approach has the potential to offer a unified treatment of all three \W\ cases, a treatment blind to the spacetime location of the C measurement (which, as noted earlier, makes no difference to the correlations). Without a prohibition on retrocausality, there seems to be no reason for any model of the ontology of these cases to care about the relative time ordering of the A, B and C measurements. Such a unified treatment would be an advantage, at least judged by the principle of not postulating ontological differences that are not reflected in experimental data. 

Our second comment  is that it is an interesting issue what happens to the case for causality from A to B, across the \W, if the three \W\ cases are unified in this way. Does the unified treatment go on the side of ED, allowing causal influence across the \W? Or does it go with DD, proposing that the Bell correlations are a selection artifact made possible by the central measurement? 

It might be thought that retrocausality will allow loophole-avoiding influence from A to B, exploiting a double  zig zag via C. But it is not clear that this is so. The other possibility is that we get zig zag dependence from A to C and B to C, but that C is still `post-selecting' in the same way. Given retrocausality,  the zig zag paths from A to C and from B to C allow Alice's and Bob's measurement choices to influence C in ED just as in DD, thus reopening the threat of counterfactual fragility. This would push all cases into the same category as DD. 

We argued in \S3 that in the case of DD, the existence of Bell correlations between A and B is very plausibly diagnosed as an artifact of conditioning on a collider at C. On standard assumptions, it is uncontroversial that C \textit{is} a collider in this case, being influenced by measurement choices both at A and at B. And our two causal tests, NDA and CFA, gave us strong reason to suspect that the AB correlation is a selection artifact, rather than a manifestation of real causal influence.

At this point, admitting retrocausality seems to cut both ways. On the one hand, it makes C a collider for influences from A and B in all three \W\ geometries. In that sense, it puts on the table the possibility that the observed Bell correlations between A and B are the result of conditioning on a collider, in all three cases. On the other hand, it weakens the case that previously existed in DD for insisting on this diagnosis. That case -- our argument for SA in DD -- turned on the assumption that C could make no difference to A and B. With retrocausality, however, that assumption can no longer be made. A diagnosis of collider bias hence seems \textit{possible,} though not \textit{mandatory} in the way that it seemed to be in DD (without retrocausality). 

Retrocausal models thus \textit{suggest,} though may not \textit{require,} the truth of the following conjecture:
\begin{description}
\item[No \W\ Causality Conjecture (NoWCC)] There is no causality across the central measurement in \W-shaped Bell experiments.  Bell correlations across such configurations are always a manifestation of  collider bias. 
\end{description}
For our third observation, we turn to some issues about the consequences and ramifications of NoWCC, especially for the retrocausal proposal itself.

 \subsection{Retrocausality and quantum colliders}

In this case we want to make a cluster of points about the consequences and ramifications of NoWCC, and it will be convenient to separate them into three. 

 \subsubsection{NoWCC as a response to two objections to retrocausal models}
First, we note that NoWCC would be a helpful result for the retrocausal approach, offering a response to two objections. As we have seen, retrocausalists propose that the spacelike nonlocal influences revealed in EPR-Bell phenomena be decomposed into zig zag influences via the intersection of the world lines of the particles involved, within their past light cones. One objection to this proposal is that the world lines of entangled particles need not have intersected in this way -- the particles may never have met, so to speak.\footnote{We first heard this objection from  a referee for  \textit{Journal of Philosophy,} in the early 1990s.} 
Another objection is that these zig zags could be chained together, allowing causation to iterate without bounds.\footnote{Travis Norsen makes this objection  \cite{Norsen15}.} 
But if NoWCC holds, then both objections fail. There is no causality to explain in cases of entanglement swapping between particles without a common past, dealing with the first objection. And causality doesn't iterate across such measurements, dealing with the second.

 \subsubsection{NoWCC as a new objection to retrocausal models?}
Second, we want to note the possibility that the line of reasoning that leads to NoWCC might be turned against the retrocausal proposal. By retrocausalist lights, the event at the vertex of an ordinary \V-shaped Bell  experiment should presumably count as a collider. After all, the retrocausalist claims that it is influenced by both of the later measurements on the two arms of the \V. Doesn't this mean that any claimed zig zag causal influence through this vertex is itself an artifact of conditioning on a collider, at least by retrocausalist lights? If successful, this objection would show that the retrocausalist proposal was self-undermining.\footnote{Do retrocausal models also have to deal with colliders at the `future' vertices A and B, once we make explicit the causal arrows controlling the settings of those measurements? No, because the settings are treated as exogenous variables, with no causes within the model, so they are not colliders. It doesn't change this conclusion if we introduce into the model a causal pathway between an earlier \textit{choice} of setting and a later \textit{implementation} of that setting. The latter would have no causes within the model other than the choice of setting, so again, it would not be a collider.}

This is a helpful objection, in part because it suggests a refinement of NoWCC. In our view, the crucial points are (i) that an experimenter's external control of a past-vertex collider is normally quite different from that of a future-vertex collider, and (ii) that this difference allows a causal channel in the first case but not the second.  

Concerning (i), there is a well-known difference between control of preparations and control of measurements. Even when a measurement looks exactly like a time-reverse of a given preparation, we lack control of the outcome of the measurement but can control the corresponding `income' (\cite{Hardy21}) of the preparation. This is why in case of measurement we resort to post-selection, throwing away outcomes that don't meet our criteria; whereas in the case of preparation, it is often the case that nothing needs to be thrown away. We simply produce the incomes we need. 

Concerning (ii), the claim is that this difference in control will explain why there is a causal channel through a past collider (where we do have control of incomes) but not through a future collider (where we don't have control of outcomes). We can check this by imaging cases in which we add such control to a future collider, and confirming that we then have a counterexample to NoWCC -- in other words, a causal channel across the future collider. This possibility is discussed  by \cite{PriceWeslake10}. Price and Weslake do not consider quantum cases, or use the terminology of colliders, but they do discuss the implications of a \textit{future} low entropy boundary constraint, analogous to those that normally enable control from the past. They argue that it would allow zig zag causation, via the future boundary. 

For a quantum case, consider quantum teleportation, which has a /\textbackslash/-shaped geometry.  The input qubit (lower left) cannot be transmitted to the output qubit (upper right) without knowing the result of the joint measurement at the future vertex.  (Depending on this outcome, a corresponding unitary transformation of the output qubit can complete the usual teleportation procedure.)  But it is easy to see that if we had control of the outcome of the measurement at the future vertex, fixing it to a particular value, a causal channel between the lower left and the upper right would be open -- the teleportation procedure could be guaranteed to occur.  

Indeed, this is precisely the scheme proposed by \cite{HorowitzMaldacena04} for solving the black hole information problem. Again, this proposal involves a /\textbackslash/-shaped geometry. 
The left input is the information being thrown into the black hole, and the bottom right vertex is a pair of entangled particles created at the event horizon.  One of these particles falls into the black hole and meets the in-falling information at some future collider inside the black hole (the upper left vertex). But this collider is such that some particular result is guaranteed to occur, due to a boundary condition at the singularity.  The net effect is that the other entangled particle that leaves the black hole on the far right encapsulates the in-falling information, so that no net information is lost. 
Horowitz and Maldacena describe the proposal as follows:  
\begin{quote}
In the process of black hole evaporation, particles are created in correlated pairs with one
falling into the black hole and the other radiated to infinity. The correlations remain even
when the particles are widely separated. The final state boundary condition at the black
hole singularity acts like a measurement that collapses the state into one associated with
the infalling matter. This transfers the information to the outgoing Hawking radiation in
a process similar to ``quantum teleportation''. \cite{HorowitzMaldacena04}
\end{quote}

\begin{figure}[t!]
 \centering
\includegraphics[height=5cm]{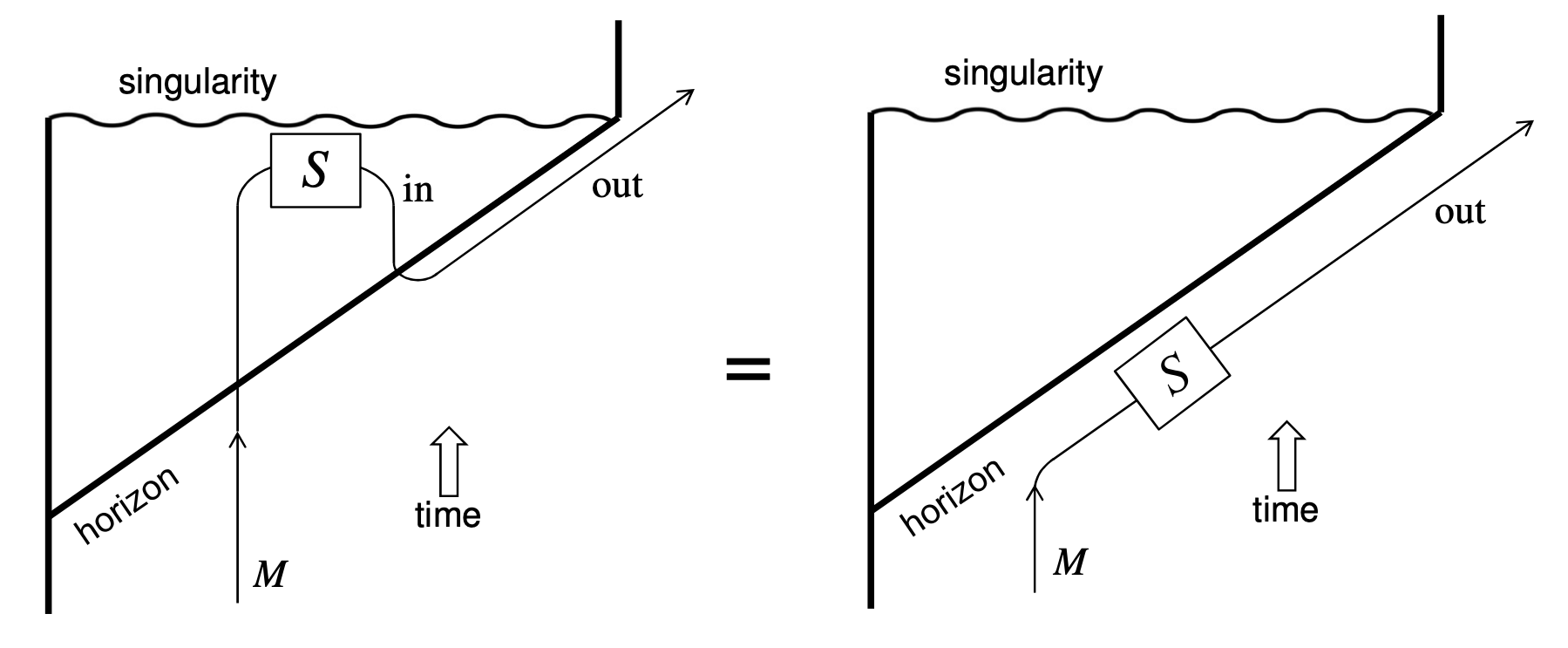}
\caption{\centering Causality via a future collider (from \cite{LloydPreskill14})?}
\end{figure}

Figure 7 is a representation of the Horowitz-Maldacena proposal by Lloyd and Preskill, whose caption for the diagram is this: 
\begin{quote}
The Horowitz-Maldacena model, in which quantum information carried by the collapsing matter system M is teleported out of a black hole. Outgoing Hawking radiation is maximally entangled with infalling radiation, and a final-state boundary condition projects M and the infalling radiation to a maximally entangled state which encodes the unitary S-matrix S.  \cite[6]{LloydPreskill14}  
\end{quote}

Similar ideas have recently been discussed by Perry \cite{Perry21a, Perry21b}, who illustrates them with Figure 8. Perry describes this figure  as the `Penrose diagram of a black hole that evaporates completely together with a river showing the expected flow of quantum information.'
According to such a proposal, Perry concludes,
\begin{quote}
 [t]he interior of the black hole is
therefore a strange place where one's classical notions of
causality and unitarity are violated. This does not matter as long as outside the black hole such pathologies do
not bother us. \cite[4]{Perry21a}

\end{quote}

For our purposes, what matters is that such proposals  induce a causal influence (Perry's river) from the in-falling matter on the lower left to the outgoing state on the upper right,  via the zig zag at the future boundary condition. This would be a counterexample to NoWCC, but one that depended on a special future boundary condition at the singularity.

\begin{figure}[t!]
 \centering
\includegraphics[height=9.5cm]{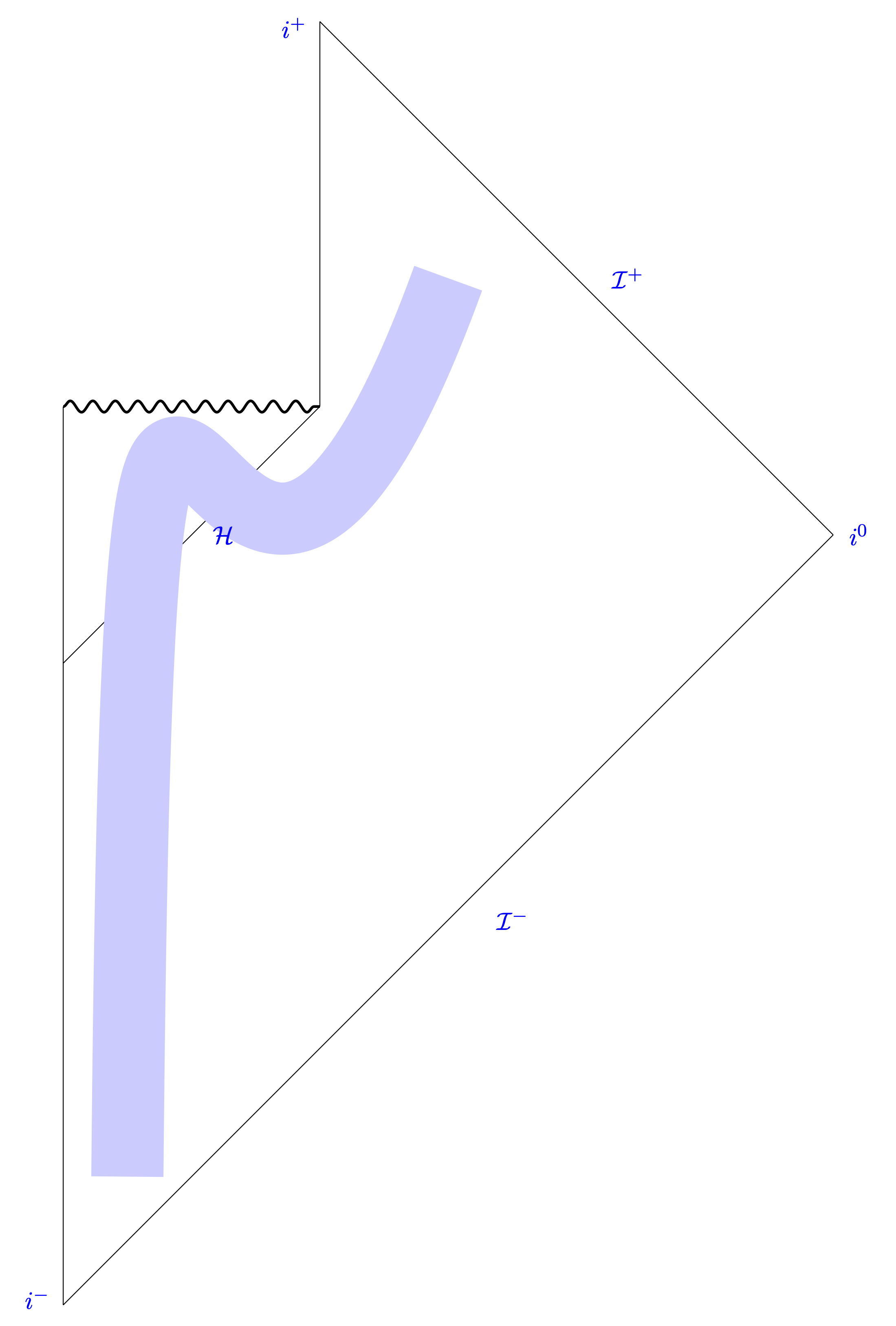}
\caption{\centering  Perry's river of quantum information (from \cite[31]{Perry21b})}
\end{figure}

These cases suggest the following improvement to NoWCC, which we now propose as an explicitly time-symmetric principle. 
\begin{description}
\item[Trans-Collider Causality Conjecture (TraCCC)] There is no causal influence across quantum colliders, except where it is enabled by external control of in/outcomes at the measurement comprising the collider. 
\end{description}
By retrocausal lights, the apparent time-asymmetry stems from the fact that control of colliders is common from the past, but rare (if not unknown) from the future.

 \subsubsection{What does TraCCC tell us about Early Deflt?}

Finally, we turn to a puzzle about the implications of NoWCC and TraCCC in ED.  In this case, the entanglement swapping measurement C occurs in the absolute past of the measurements at A and B. In principle, its result could be transmitted to Alice and Bob before they choose the settings for the latter measurements. In other words, they may know that that a given instance falls within $E_{C}$, before they choose the settings $a_n$ and $b_n$. There is post-selection, but it has happened in the past, from Alice and Bob's point of view. Doesn't this amount to control of the outcome at C? If so, then TraCCC will allow a causal channel through C in ED, contrary to our proposal that retrocausal models treat the three \W\ cases in the same way. 

The reasoning that led us to CFA suggests an answer here. In the retrocausal picture, applied to ED, Alice may indeed know that whichever $a_n$ she chooses, what results is a given C measurement outcome, $C_n$, known to her in advance. Before she chooses $a_n$ -- i.e., at the point at which she doesn't yet know what she will choose -- she is right to say:
\begin{description}
\item[Before] Whichever setting I choose, the result at C will be (is)  $C_n$. 
\end{description}
Does it follow that \textit{after} she has chosen, she is right to say the following?
\begin{description}
\item[After] If I \textit{had} chosen the alternative setting, the result would still have been $C_n$
\end{description}
 A  retrocausalist who wants to claim  that Alice's choice may influence C has to deny that Alice is entitled to \textbf{After}. To admit the possibility of retrocausality from A to C -- i.e., to allow that Alice's choice at A might \textit{make a difference} at C -- just {is} to claim that this counterfactual need not hold. 
 
 So long as \textbf{After} need not hold,  $E_{C}$ is not counterfactually robust, and the claim of causality across the collider falls to counterfactual fragility test (CFA). Contrast this to the case in which we imagine the C outcome under the control of a demon, Charlie, who simply \textit{ensures} that it takes the value $C_n$. In this case, Alice would be entitled to the counterfactual \textbf{After}, and the claim of causality across the \W\ would avoid CFA.

Still, it is one thing to observe that this is the line that the retrocausalists need to take in such a case, if they are to invoke TraCCC to deny causality across the \W\ in ED. It is another to show that they can get away with it. We flag this an issue for further work, and close with two observations. 

First, retrocausalists face a similar issue in (some) ordinary \V-shaped experiments. If measurement B lies in the past light cone of  measurement A, then Alice may know the outcome of B, before she chooses her own setting. If she is a retrocausalist, this will put her in a similar position, affirming an analogue of \textbf{Before}, but denying an analogue of \textbf{After}.
Second, the key to the consistency of this combination, if there is one, presumably rests on a sense in which \textbf{Before} and \textbf{After} `look in different directions'. In possible world terms, \textbf{Before} is a claim about the actual world. As a counterfactual, however, \textbf{After} looks to a non-actual world -- one in which Alice's choice differs from its actual value.
\end{appendices}


\begin{thebibliography}{999}

\bibitem[Bacciagaluppi \& Hermens 2021]{Guido21}
Bacciagaluppi, G.~\& Hermens, R. Bell Inequality Violation and Relativity of Pre- and Postselection. arXiv:2002.03935



\bibitem[Bell 1979]{Bell79}
Bell, J.~S. Atomic-cascade photons and quantum-mechanical nonlocality. Reprinted in \cite[105--110]{Bell04}. 
\bibitem[Bell 1990]{Bell90}
Bell, J.~S. La nouvelle cuisine. In \textit{Between Science and Technology,} edited by A. Sarlemijn and P. Kroes (Elsevier Science, Amsterdam). Reprinted in \cite[232--248]{Bell04}.



\bibitem[Bell 2004]{Bell04}
Bell, J.~S. \textit{Speakable and Unspeakable in Quantum Mechanics, Second Edition.} Cambridge University Press: Cambridge.

\bibitem[Berkson 1946]{Berkson46}
Berkson, J. Limitations of the application of fourfold table analysis to hospital data. \textit{Biometrics Bulletin,}  2,  47--53.


\bibitem[Blasiak, Borsuk \& Markiewicz 2020]{Blasiak20}
Blasiak, P., Borsuk, E. \& Markiewicz, M. On safe post-selection for Bell nonlocality: Causal diagram approach. arXiv:2012.07285 

\bibitem[Clifton et al 1990]{Clifton90}
Clifton, R., Butterfield, J. and Redhead, M., Nonlocal influences and possible worlds---A Stapp in the wrong direction, \textit{British ]ournal for the Philosophy of Science,} 41, 5--58.

\bibitem[Cole et al 2010]{Cole10}
Cole, S', Platt, R., Schisterman, E., Chu, H., Westreich, D., Richardson, D., Poole, C. Illustrating bias due to conditioning on a collider. \textit{International Journal of Epidemiology,} 39,  417--420. https://doi.org/10.1093/ije/dyp334

\bibitem[Egg 2013]{Egg13}
Egg, M. Delayed-choice experiments and the metaphysics of entanglement.
\textit{Foundations of Physics,} 43, 1124--1135.


\bibitem[Fankhauser 2019]{Fankhauser19}
Fankhauser, J. Taming the delayed choice quantum eraser. \textit{Quanta,} 8, 44--56. arXiv:1707.07884
 


\bibitem[Friedrich \& Evans 2019]{FriedrichEvans19}
Friederich, S. \& Evans, P. Retrocausality in quantum mechanics. In \textit{The Stanford Encyclopedia of Philosophy} (Summer 2019 Edition); Zalta, E., Ed. \href{http://plato.stanford.edu/archives/sum2019/entries/qm-retrocausality/}{http://plato.stanford.edu/archives/sum2019/entries/qm-retrocausality/}

\bibitem[Gaasbeek 2010]{Gaasbeek10}
Gaasbeek, B. Demystifying the delayed choice experiments. arXiv:1007.3977

\bibitem[Glick 2019]{Glick19}
Glick, D. Timelike entanglement for delayed-choice entanglement swapping, \textit{Studies in History and Philosophy of Modern Physics,} 68, 16--22.

\bibitem[Giustina et al 2015]{Giustina15}
Giustina, M., Versteegh, M.A., Wengerowsky, S., Handsteiner, J., Hochrainer, A., Phelan, K., et al. Significant-loophole-free test of Bell’s theorem with entangled photons. \textit{Physical review letters,} 115(25), 250401. arxiv:1511.03190

\bibitem[Hardy 2021]{Hardy21}
Hardy, L. Time symmetry in operational theories. arXiv:2104.00071

\bibitem[Healey 2012]{Healey12}
Healey, R. Quantum theory: a pragmatist approach. \textit{The British Journal for the Philosophy of Science,} 63, 729--771.

\bibitem[Hensen et al 2015]{Hensen15}
Hensen, B., Bernien, H., Dreau, A. E., Reiserer, A., Kalb, N., Blok, M. S., et al. Loophole-free Bell inequality violation using electron spins separated by 1.3 kilometres. \textit{Nature,} 526, 682--686. arXiv:1508.05949

\bibitem[Horowitz \& Maldacena 2004]{HorowitzMaldacena04}
Horowitz, G.~and Maldacena, J. The black hole final state, \textit{JHEP} 0402:008 (2004). arXiv:hep-th/0310281

\bibitem[Lloyd \& Preskill 2014]{LloydPreskill14}
Lloyd, S \& Preskill, J. Unitarity of black hole evaporation in final-state projection models, \textit{Journal of High Energy Physics,} 2014, 126. arXiv:1308.4209

\bibitem[Ma et al 2012]{Ma12}
Ma, X.-s., Zotter, S., Kofler, J., Ursin, R., Jennewein, T., Brukner, C., et al.
Experimental delayed-choice entanglement swapping. \textit{Nature Physics,} 8,
479--484. arXiv:1203.4834

\bibitem[Merali 2015]{Merali15}
Merali, Z., Quantum `spookiness' passes toughest test yet. \emph{Nature,} 27 August 2015 [\href{http://www.nature.com/news/quantum-spookiness-passes-toughest-test-yet-1.18255}{http://www.nature.com/news/quantum-spookiness-passes-toughest-test-yet-1.18255}].

\bibitem[Norsen 2011]{Norsen11}
Norsen, T. John S.~Bell’s concept of local causality.
\textit{American Journal of Physics} 79, 1261. https://doi.org/10.1119/1.3630940

\bibitem[Norsen 2015]{Norsen15}
Norsen, T. Reply To: Retrocausality is intrinsic to quantum mechanics. \textit{International Journal of Quantum Foundations.} July 17, 2015. \href{https://ijqf.org/forums/reply/2832}{https://ijqf.org/forums/reply/2832}

\bibitem[Norsen \& Price 2021]{NorsenPrice21}
Norsen, T. \& Price, H. Lapsing quickly into fatalism: Bell on backward causation. \textit{Entropy,} 23(2021), 251.

\bibitem[Peres 2000]{Peres00}
Peres, A. Delayed choice for entanglement swapping. \textit{Journal of Modern Optics,} 47, 139--143.

\bibitem[Perry 2021a]{Perry21a}
Perry, M. No Future in Black Holes. arXiv:2106.03715.

\bibitem[Perry 2021b]{Perry21b}
Perry, M. Future Boundaries and the Black Hole Information Paradox. arXiv:2108.05744.




\bibitem[Price \& Weslake 2010]{PriceWeslake10}
Price, H.~\& Weslake, B. The time-asymmetry of causation. In Helen Beebee, Christopher Hitchcock and Peter Menzies (eds), \textit{The Oxford Handbook of Causation} (OUP), 414--443.

\bibitem[Rosenfeld et al 2017]{Rosenfeld17}
Rosenfeld, W., Burchardt, D., Garthoff, R., Redeker, K., Ortegel, N., Rau, M., \& Weinfurter, H. Event-ready Bell test using entangled atoms simultaneously closing detection and locality loopholes. \textit{Phys.~Rev.~Lett.} 119, 010402. arXiv:1611.04604

\bibitem[Shalm et al 2015]{Shalm15}
Shalm, L.K., Meyer-Scott, E., Christensen, B.G., Bierhorst, P., Wayne, M.A., Stevens, M.J., et al. Strong loophole-free test of local realism. \textit{Physical Review Letters,} 115(25), 250402. arxiv:1511.03189

\bibitem[Wharton \& Argaman 2020]{WhartonArgaman20}
Wharton, K. \& Argaman, N. Bell's Theorem and locally-mediated reformulations of quantum mechanics. \emph{Reviews of Modern Physics,} \textit{92,} 21002. arXiv:1906.04313.
\end{thebibliography}
\end{document}